\documentclass[]{article}

\let\realleq\leq

\usepackage{graphicx}
\usepackage{amssymb}

\newtheorem{theorem}{Theorem}

\newtheorem{corollary}[theorem]{Corollary}

\newtheorem{definition}[theorem]{Definition}

\newtheorem{proposition}[theorem]{Proposition}

\newenvironment{proof}[1][Proof]{\noindent\textbf{#1.} }{\ \rule{0.5em}{0.5em}}

\newcommand\overset{\stackrel}
\newcommand\MPCPS{\textit{Math. Proc. Camb. Phil. Soc.} }
\newcommand\JMP{\textit{J. Math. Phys.} }
\newcommand\IJTP{\textit{Int. J. Theor. Phys.} }
\newcommand\PR{\textit{Phys. Rev.} }

\newcommand\GRG{\textit{Gen. Rel. Grav.} }

\def\sep{\unskip, }                  
\def\MSC{\par\leavevmode\hbox {\it 1991 MSC:\ }}%
\def\PACS{\par\leavevmode\hbox {\it PACS:\ }}%

\begin{document}

\title{Lightlike simultaneity, comoving observers and distances in general relativity}

\author{V. J. Bol\'os \\{\small Dpto. Matem\'aticas, Facultad de Ciencias, Universidad de
Extremadura.}\\ {\small Avda. de Elvas s/n. 06071--Badajoz,
Spain.}\\ {\small e-mail\textup{: \texttt{vjbolos@unex.es}}}}

\maketitle

\PACS 04.20.Cv \sep 02.40.Hw \MSC 53A35 \sep 53B30 \sep 83C99

\begin{abstract}
We state a condition for an observer to be comoving with another
observer in general relativity, based on the concept of lightlike
simultaneity. Taking into account this condition, we study
relative velocities, Doppler effect and light aberration. We
obtain that comoving observers observe the same light ray with the
same frequency and direction, and so gravitational redshift effect
is a particular case of Doppler effect. We also define a distance
between an observer and the events that it observes, that
coincides with the known \textit{affine distance}. We show that
affine distance is a particular case of radar distance in the
Minkowski space-time and generalizes the proper radial distance in
the Schwarzschild space-time. Finally, we show that affine
distance gives us a new concept of distance in Robertson-Walker
space-times, according to Hubble law.
\end{abstract}

\section{Introduction}

In general relativity it is often difficult to interprete when an
observer $\beta $ is comoving with another observer $\beta '$, in
the sense that $\beta $ moves ``like" $\beta '$. For example,
given a particular coordinate system it is usual to suppose that
stationary observers (i.e. with constant spatial coordinates) are
comoving each one with respect to the other. But this criterion is
coordinate-dependent: let us suppose that two observers are
stationary using a particular coordinate system; then they are
comoving each one with respect to the other. On the other hand, we
can find another coordinate system in which one observer is
stationary and the other one is not stationary; then they are not
comoving each one with respect to the other. Since we want that
the property ``to be comoving with" was an intrinsic property of
the observer (i.e. that an observer was able to decide if it is
comoving with another observer or not, independently from the
coordinate system), the ``stationary criterion" is a bad
criterion.

Given an observer $\beta $, there is a general method to check if
it is comoving with another observer $\beta '$, based on the
concept of simultaneity. We have to build a simultaneity foliation
associated with $\beta $ \cite{Bolo02}, then parallelly transport
the 4-velocity of $\beta '$ to $\beta $, along geodesics joining
$\beta '$ with $\beta $ in the leaves of the foliation, and
finally compare it with the 4-velocity of $ \beta $ (see Fig.
\ref{figintro}).

There are a lot of kinds of simultaneities, but we are going to
consider only two kinds of simultaneity foliations associated with
a given observer $\beta $ \cite{Bolo02}: the Landau foliation
$\mathcal{L}_{\beta }$, whose leaves are Landau submanifolds
\cite{Oliv80}, also called Fermi surfaces (spacelike); and the
past-pointing horismos foliation $\mathcal{E}_{\beta }^{-}$, whose
leaves are past-pointing horismos submanifolds \cite{Beem81}
(lightlike). We have to note that if we use Landau foliations,
then the method to check if an observer is comoving with another
one is symmetric; on the other hand, if we use past-pointing
horismos foliations, then this method is not symmetric, i.e. one
observer $\beta $ can be comoving with another observer $\beta '$,
and $\beta '$ being non comoving with $\beta $. But, since we are
working in relativity, the non-symmetry is an acceptable property.
So, the problem is to decide which simultaneity (spacelike or
lightlike) is mathematically and physically more suitable for us:

\begin{figure}[tbp]
\begin{center}
\includegraphics[width=0.65\textwidth]{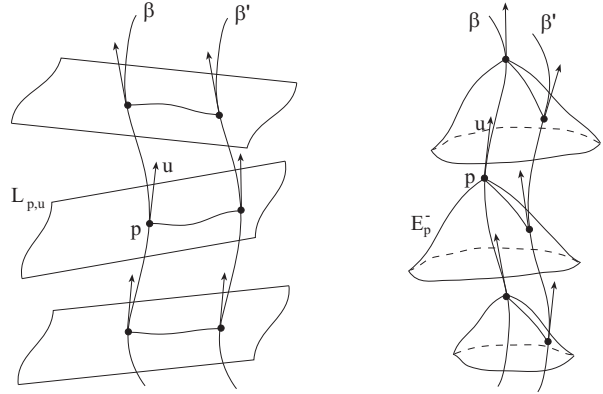}
\end{center}
\caption{How to check if an observer $\beta $ is comoving with
another observer $\beta '$, depending on the simultaneity
foliation that we are using. Left: Landau foliation
$\mathcal{L}_{\beta }$ (spacelike). Right: Past-pointing horismos
foliation $\mathcal{E}_{\beta }^{-}$ (lightlike).}
\label{figintro}
\end{figure}

\begin{description}
\item[(a)] Mathematically: in a previous work \cite{Bolo02} we
proved that the Landau foliation $\mathcal{L}_{\beta }$ is not
always defined in every convex normal neighborhood because its
leaves can intersect themselves. For example, in a Minkowski
space-time if the observer $\beta $ is not geodesic. Moreover,
$\mathcal{L}_{\beta }$ is not necessarily spacelike at every point
of a convex normal neighborhood. On the other hand, the
past-pointing horismos foliation $\mathcal{E}_{\beta }^{-}$ is
always well defined in every convex normal neighborhood and it is
always lightlike.

\item[(b)] Physically: given an observer at an event $p$ with
4-velocity $u$, the events of its Landau submanifold $L_{p,u}$ do
not affect the observer at $p$ in any way, since both
electromagnetic and gravitational waves travel at the speed of
light. On the other hand, the events of its past-pointing horismos
submanifold $E_{p}^{-}$ are precissely the events that affect and
are observed by the observer at $p$, i.e. the events that
\textit{exist} for the observer at $p$.
\end{description}

Therefore, we are going to work in the framework of lightlike
simultaneity. So, given an observer at an event $p$, we will say
that the events of $E_{p}^{-}$ are \textit{lightlike-simultaneous}
for this observer at $p$. In fact \textquotedblleft to be
lightlike-simultaneous for an observer\textquotedblright\ is the
same as \textquotedblleft to be observed simultaneously by an
observer\textquotedblright .

Hence, in Section \ref{S3}, we define the \textit{observers
congruence comoving with a given observer}, according to the
concept of lightlike simultaneity, and we give a method to measure
relative velocities of observers in Section \ref{S3.1}. Given a
light ray, we study Doppler effect in Section \ref{S3.2},
obtaining that the frequency of a light ray remains constant for
comoving observers. This is apparently contradictory with
\textit{gravitational redshift} effect, stating that light rays
gain or lose frequency in the presence of a gravitational field,
and it is considered independent of Doppler effect. Gravitational
redshift effect is completely explained in our formalism, showing
that it is a particular case of a generalized Doppler effect. We
also study light aberration effect in Section \ref{S3.3},
obtaining that there is not light aberration between comoving
observers.

The concept of distance is strongly bounded to the concept of
simultaneity too. We are using lightlike simultaneity, so we have
to measure distances between lightlike-simul\-ta\-neous events,
i.e. we need to measure lengths of light rays. In Section
\ref{S4}, we re-define a concept of distance (called
\textit{affine distance}) between an observer and the events that
it observes, i.e. a distance between $p$ and the events of
$E_{p}^{-}$. In Section \ref{S5}, we show that affine distance is
a particular case of radar distance in the Minkowski space-time
and generalizes the proper radial distance in the Schwarzschild
space-time. Finally, we show that affine distance gives us a new
concept of distance in Robertson-Walker space-times, according to
Hubble law.

We work in a 4-dimensional lorentzian space-time manifold
$\mathcal{M}$, with $c=1$ and $\nabla $ the Levi-Civita
connection, using the Landau-Lifshitz Spacelike Convention (LLSC).
We suppose that $\mathcal{M}$ is a convex normal neighborhood
\cite{Helg62}. Thus, given two events $p$ and $q$ in
$\mathcal{M}$, there exists a unique geodesic joining $p$ and $q$
. The parallel transport from $p$ to $q$ along this geodesic will
be denoted by $\tau _{pq}$. If $\beta :I\rightarrow \mathcal{M}$
is a curve with $I\subset \mathbb{R}$ a real interval, we will
identify $\beta $ with the image $\beta I$ (that is a subset in
$\mathcal{M}$), in order to simplify the notation. If $u$ is a
vector, then $u^{\bot }$ denotes the orthogonal space of $u$.
Moreover, if $x$ is a spacelike vector, then $\Vert x\Vert $
denotes the module of $x$. Given a pair of vectors $u,v$, we use
$g\left( u,v\right) $ instead of $u^{\alpha }v_{\alpha }$. If $X$
is a vector field, $X_{p}$ will denote the unique vector of $X$ in
$T_{p}\mathcal{M}$.

\section{Preliminaries}

\label{S1.1}

An \textit{observer} in the space-time is determined by a timelike
world line $\beta $, and the events of $\beta $ are the
\textit{positions} of the observer. It is usual to identify an
observer with its world line, and so $\beta $ is an observer. The
\textit{4-velocity} of the observer is a future-pointing timelike
unit vector field $U$ defined in $\beta $ and tangent to $\beta $.
Given an event $p$, the 4-velocity of an observer at $p$ is given
by a future-pointing timelike unit vector $u$. It is also usual to
identify an observer with its 4-velocity, since they are defined
reciprocally. So, if $u$ is the 4-velocity of an observer at $p$,
we will say that $u$ is an observer at $p$, in order to simplify
the notation. To sum up, we will say that a timelike world line
$\beta $ is an observer, and a future-pointing timelike unit
vector $u$ in $T_p\mathcal{M}$ is an observer at $p$.

Given two observers $u$ and $u'$ at the same event $p$, there
exists a unique vector $v\in u^{\bot }$ and a unique positive real
number $\gamma $ such that
\begin{equation}
u'=\gamma \left( u+v\right) .  \label{f1.uprima}
\end{equation}
As consequences, we have $0\realleq\Vert v\Vert <1$ and $\gamma
=-g\left( u',u\right) =\frac{1}{\sqrt{1-\Vert v\Vert ^{2}}}$. We
will say that $v$ is the \textit{relative velocity of
}$u'$\textit{\ observed by }$u$, and $\gamma $ is the
\textit{gamma factor} corresponding to the velocity $\Vert v\Vert
$.


A \textit{light ray} is given by a lightlike geodesic $\lambda $
and a future-pointing lightlike vector field $F$ defined in
$\lambda $, tangent to $\lambda $ and parallelly transported along
$\lambda $ (i.e. $\nabla _{F}F=0$), called \textit{frequency
vector field of }$\lambda $. Given $p\in \lambda $ and $u$ an
observer at $p$, there exists a unique vector $w\in u^{\bot }$ and
a unique positive real number $\nu $ such that
\begin{equation}
F_{p}=\nu \left( u+w\right) .  \label{f1.efepe}
\end{equation}
As consequences, we have $\Vert w\Vert =1$ and $\nu =-g\left(
F_{p},u\right) $. We will say that $w$ is the \textit{relative
velocity of }$\lambda $ \textit{\ observed by }$u$, and $\nu $ is
the \textit{frequency of }$\lambda $\textit{\ observed by }$u$. In
other words, $\nu $ is the module of the projection of $F_{p}$
onto $u^{\perp }$.

A \textit{light ray from }$q$\textit{\ to }$p$ is a light ray
$\lambda $ such that $q$, $p\in \lambda $ and $\exp _{q}^{-1}p$ is
future-pointing.

Given two observers $u$ and $u'$ at the same event $p$ of a light
ray $\lambda $, using (\ref{f1.efepe}), the frequency vector
$F_{p}$ of $\lambda $ is given by
\begin{equation}
F_{p}=\nu \left( u+w\right) =\nu '\left( u'+w'\right) ,
\label{f1.proporcio}
\end{equation}
where $\nu $, $\nu '$ are the frequencies of $\lambda $ observed
by $u$, $u'$ respectively and $w$, $w'$ are the relative
velocities of $\lambda $ observed by $u$, $u'$ respectively.
Applying (\ref{f1.uprima}), we obtain that
\begin{equation}
\nu '=\gamma \left( 1-g\left( v,w\right) \right) \nu .
\label{f1.dop}
\end{equation}
Expression (\ref{f1.dop}) is the general expression of
\textit{Doppler effect}. For example, if $\frac{v}{\Vert v\Vert
}=w$, i.e. the direction of the relative velocity of $u'$ observed
by $u$ coincides with the direction of the relative velocity of
$\lambda $ observed by $u$, we have the usual redshift expression
$\nu '=\sqrt{\frac{1-\Vert v\Vert }{1+\Vert v\Vert }}\nu $.

On the other hand, taking into account (\ref{f1.proporcio}) and
(\ref{f1.dop}), we have
\begin{equation}
w'=\frac{1}{\gamma \left( 1-g\left( v,w\right) \right) } \left(
u+w\right) -u'.  \label{f1.propor}
\end{equation}
The fact that $w'$ is different from $w$ causes an
\textit{aberration} effect \cite{Bolo05}. It is easy to prove that
\begin{equation}
\cos \theta =\frac{\cos \theta '-\Vert v\Vert }{1-\Vert v\Vert
\cos \theta '},  \label{f1.cosabe}
\end{equation}
where $\theta $ is the angle between $-w$ and $v$, and $\theta '$
is the angle between $-w'$ and the projection of $v$ onto
$u^{\prime \bot }$ ($\theta '$ is also the angle between $-w'$ and
$-v'$, where $v'$ is the relative velocity of $u$ observed by
$u'$). The expression (\ref{f1.cosabe}) is the general expression
of light aberration phenomenon \cite{Syng65}, and the scalar
function given by $ \theta '-\theta $ is the \textit{aberration
angle of }$u'$\textit{\ observed by }$u$\textit{\ corresponding
to} $\lambda $.

Let $p\in \mathcal{M}$ and $\varphi :\mathcal{M}\rightarrow
\mathbb{R}$ defined by $\varphi \left( q\right) :=g\left( \exp
_{p}^{-1}q,\exp _{p}^{-1}q\right) $. Then, it is a submersion and
the set
\begin{equation}
E_{p}:=\varphi ^{-1}\left( 0\right) -\left\{ p\right\}
\label{f1.horismos}
\end{equation}
is a regular 3-dimensional submanifold, called \textit{horismos
submanifold of }$p$ \cite{Beem81}. In other words, an event $q$ in
the space-time is in $E_{p}$ if and only if $q\neq p$ and there
exists a lightlike geodesic joining $p$ and $q$. $E_{p}$ has two
connected components, $E_{p}^{+}$ and $E_{p}^{-}$ \cite{Sach77};
$E_{p}^{+}$ (respectively $E_{p}^{-}$) is the
\textit{future-pointing} (respectively \textit{past-pointing})
\textit{horismos submanifold of }$p$, and it is the connected
component of (\ref{f1.horismos}) in which, for each event $q\in
E_{p}^{+}$ (respectively $q\in E_{p}^{-}$), the preimage $\exp
_{p}^{-1}q$ is a future-pointing (respectively past-pointing)
lightlike vector.

We can construct horismos foliations in this way (\cite{Bolo02},
\cite{Bolo03}): let $\beta $ be an observer. Then, we define
$\mathcal{M}_{\beta }^{+}:=\cup _{p\in \beta }E_{p}^{+}$ and
$\mathcal{M}_{\beta }^{-}:=\cup _{p\in \beta }E_{p}^{-}$. So,
there exists a foliation $\mathcal{E}_{\beta }^{+}$ (respectively
$\mathcal{E}_{\beta }^{-}$) defined in $\mathcal{M}_{\beta }^{+}$
(respectively $\mathcal{M}_{\beta }^{-}$) whose leaves are
future-pointing (respectively past-pointing) horismos submanifolds
of events of $\beta $. The foliations $\mathcal{E}_{\beta }^{+}$
and $\mathcal{E}_{\beta }^{-}$ are called respectively
\textit{future-pointing} and \textit{past-pointing horismos
foliation generated by }$\beta $.

\section{Comoving observers in the framework of lightlike simultaneity}

\label{S3}

As we discussed in the Introduction, we are going to work in the
framework of lightlike simultaneity. So, to check if an observer
$\beta $ \textit{is comoving with} another observer $\beta '$, we
have to parallelly transport the 4-velocity of $\beta '$ to $\beta
$, along lightlike geodesics joining $\beta '$ with $\beta $ in
the leaves of the foliation $\mathcal{E}_{\beta }^{-}$, and
finally compare it with the 4-velocity of $\beta $ (see Fig.
\ref{figintro}-right). This is a non-symmetric method, i.e. if
$\beta $ is comoving with $\beta '$ then $\beta '$ is not
necessarily comoving with $\beta $.

\begin{figure}[tbp]
\begin{center}
\includegraphics[width=0.35\textwidth]{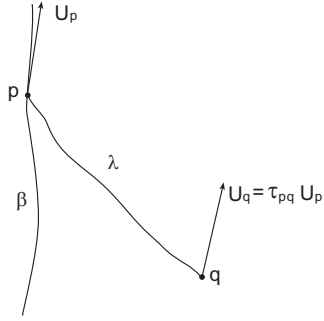}
\end{center}
\caption{The extension of $U$ at $q$ is given by $\tau
_{pq}U_{p}$, where $p\in \beta $ and there exists a light ray
$\lambda $ from $q$ to $p$. So, we can build a reference frame
from a single observer.} \label{fig1}
\end{figure}

Given an observer $\beta $ with 4-velocity $U$, we can construct
an observers congruence extending $U$ to $\mathcal{M}_{\beta
}^{-}$ by means of parallel transports along light rays from
events of $\mathcal{M}_{\beta }^{-}$ to events of $\beta $:
\begin{definition}
\label{comoving} Let $\beta $ be an observer with 4-velocity $U$.
The observers congruence associated with $\beta $ is the extension
of $U$ defined on $\mathcal{M}_{\beta }^{-}\cup \beta $ such that
$U_{q}:=\tau _{pq}U_{p}$, where $p\in \beta $, $q\in
\mathcal{M}_{\beta }^{-}$, and there exists a light ray from $q$
to $p$ (see Fig. \ref{fig1}).

Let $\beta $, $\beta '$ be two observers. We will say that $\beta
$ is comoving with $\beta '$ if $\beta '$ is an observer of the
observers congruence associated with $\beta $, i.e. $\beta '$ is
an integral curve of this vector field.
\end{definition}

Since parallel transport conserves metric and causality, the
observers congruence associated with a given observer $\beta $ is
actually an observers congruence, because it is a future-pointing
timelike unit vector field defined in the open set
$\mathcal{M}_{\beta }^{-}\cup \beta $. Moreover, $\beta $ observes
that its $4$-velocity is the same as the $4$-velocity of any
observer of this congruence. So, they define a reference frame
associated with the observer $\beta $ in a natural way.

According to this method, we state the next definition:

\begin{definition}
\label{ucomoving} Let $\lambda $ be a light ray from $q$ to $p$
and let $u$, $u'$ be two observers at $p$, $q$ respectively. We
will say that $u$ is comoving with $u'$ if $\tau _{qp}u'=u$.
\end{definition}

\subsection{Relative velocity of an observer}

\label{S3.1}

We can generalize the concept of \textquotedblleft relative
velocity of an observer\textquotedblright\ (given in Section
\ref{S1.1}) for observers at two different events of the same
light ray:

\begin{definition}
\label{relvelu} Let $\lambda $ be a light ray from $q$ to $p$ and
let $u$, $u'$ be two observers at $p$, $q$ respectively. The
relative velocity of $u'$ observed by $u$ is the relative velocity
of $\tau _{qp}u'$ observed by $u$.
\end{definition}

So, the relative velocity of $u'$ observed by $u$ is given by the
unique vector $v\in u^{\bot }$ such that $\tau _{qp}u'=\gamma
\left( u+v\right) $, where $\gamma $ is the gamma factor
corresponding to the velocity $\Vert v\Vert $. Note that $\tau
_{qp}u'$ is the way $u $ observers $u'$, and so, it is the natural
adaptation of $u'$ at $p$.

We can generalize this definition for two observers $\beta $ and
$\beta '$:

\begin{definition}
\label{relvelbeta} Let $\beta $, $\beta '$ be two observers, and
let $U$, $U'$ be the $4$-velocities of $\beta $, $\beta '$
respectively. The relative velocity of $\beta '$ observed by
$\beta $ is a vector field $V$ defined on $\beta $ such that $V_p$
is the relative velocity of $U'_q$ observed by $U_p$ (in the sense
of Definition \ref{relvelu}), where $p$, $q$ are events of $\beta
$, $\beta '$ respectively and there exists a light ray from $q$ to
$p$.
\end{definition}

By Definitions \ref{ucomoving} and \ref{relvelu}, we have that $u$
is comoving with $u'$ if and only if the relative velocity of $u'$
observed by $u$ is zero. Analogously, by Definitions
\ref{comoving} and \ref{relvelbeta}, we have that $\beta $ is
comoving with $\beta '$ if and only if the relative velocity of
$\beta '$ observed by $\beta $ is zero.

For example, in the Schwarzschild metric with spherical
coordinates $ds^{2}=-a^{2}\left( r\right)
dt^{2}+\frac{1}{a^{2}\left( r\right) }dr^{2}+r^{2} \left( d\theta
^{2}+\sin ^{2}\theta d\varphi ^{2}\right) $, where $a\left(
r\right) =\sqrt{1-\frac{2m}{r}}$ and $r>2m$, we have that $\lambda
:\left[ r_1,+\infty \right) \rightarrow \mathcal{M}$ with $r_1>2m$
given by
\begin{equation}
\label{lightray} \lambda \left( r\right) :=\left( 2m\ln \left(
\frac{r-2m}{r_1-2m} \right) +r-r_1,r,\frac{\pi }{2},0\right)
\end{equation}
is a radial light ray emitted from $q:=\lambda \left( r_1\right)
=\left( 0,r_1,\pi /2,0\right) $ and moving away from the event
horizon $r=2m$. Given a radius $r_2>r_1$, let $p:=\lambda \left(
r_2\right) $ be an event of $\lambda $ and let $u_1=u_1^t\left.
\frac{\partial }{\partial t}\right\vert _{q}+u_1^r\left.
\frac{\partial }{\partial r}\right\vert _{q}+u_1^{\theta }\left.
\frac{\partial }{\partial \theta }\right\vert _{q}+u_1^{\varphi
}\left. \frac{\partial }{\partial \varphi }\right\vert _{q}$ be a
vector in $T_q\mathcal{M}$. Taking into account the Christoffel
symbols of the metric, it can be proved that
\begin{eqnarray}
\tau _{qp}u_1&=&\frac{1}{2a_2^2}\left( \left( a_2^2+a_1^2\right)
u_1^t+\left( 1-\frac{a_2^2}{a_1^2}\right) u_1^r\right) \left.
\frac{\partial }{\partial t}\right\vert _{p} \nonumber
\\
& &+\frac{1}{2}\left( \left( a_1^2-a_2^2\right) u_1^t+\left(
1+\frac{a_2^2}{a_1^2}\right) u_1^r\right) \left. \frac{\partial
}{\partial r}\right\vert _{p} \nonumber
\\
& &+\frac{r_1}{r_2}u_1^{\theta }\left. \frac{\partial }{\partial
\theta }\right\vert _{p}+\frac{r_1}{r_2}u_1^{\varphi }\left.
\frac{\partial }{\partial \varphi }\right\vert _{p} ,
\label{tauqp}
\end{eqnarray}
where $a_1:=a\left( r_1\right) $ and $a_2:=a\left( r_2\right) $.

\begin{itemize}
\item If $u_1$ is a stationary observer, then
$u_1=\frac{1}{a_1}\left. \frac{\partial }{\partial t}\right\vert
_{q} $. So, by (\ref{tauqp}), we have $\tau
_{qp}u_1=\frac{1}{2a_1}\left( \left( 1+\frac{a_1^2}{a_2^2}\right)
\left. \frac{\partial }{\partial t}\right\vert _{p}+\left(
a_1^2-a_2^2\right) \left. \frac{\partial }{\partial r}\right\vert
_{p}\right) $. Let $u_2$ be a stationary observer at $p$. Taking
into account Definition \ref{relvelu}, the relative velocity $v$
of $u_1$ observed by $u_2$ is given by
\begin{equation}
\label{relvelschw} v=a_2\frac{a_1^2-a_2^2}{a_1^2+a_2^2}\left.
\frac{\partial }{\partial r}\right\vert _{p} ,
\end{equation}
and hence, $\Vert v\Vert =\frac{a_2^2-a_1^2}{a_2^2+a_1^2}<1$. If
$r_1\rightarrow 2m$ then $\Vert v\Vert \rightarrow 1$. This
accords with the fact that ``a `particle' at rest in the space at
$r=2m$ would have to be a photon'' \cite{Rind79}.

\item If $u_1$ is a radial free-falling observer, then
$u_1=\frac{E}{a_1^2}\left. \frac{\partial }{\partial t}\right\vert
_{q}-\sqrt{E^2-a_1^2}\left. \frac{\partial }{\partial
r}\right\vert _{q} $, where $E$ is a constant of motion given by
$E:=\left( \frac{1-2m/r_0}{1-v_0^2}\right) ^{1/2}$, $r_0$ is the
radial coordinate at which the fall begins, and $v_0$ is the
initial velocity (see \cite{Craw02}). Let $u_2$ be a stationary
observer at $p$. So, by (\ref{tauqp}) and taking into account
Definition \ref{relvelu}, the relative velocity $v$ of $u_1$
observed by $u_2$ is given by
\[
v=-a_2\frac{\left( a_2^2+a_1^2\right) \sqrt{E^2-a_1^2}+E\left(
a_2^2-a_1^2\right) }{\left( a_2^2-a_1^2\right)
\sqrt{E^2-a_1^2}+E\left( a_2^2+a_1^2\right) }\left. \frac{\partial
}{\partial r}\right\vert _{p} ,
\]
and hence, $\Vert v\Vert =\frac{\left( a_2^2+a_1^2\right)
\sqrt{E^2-a_1^2}+E\left( a_2^2-a_1^2\right) }{\left(
a_2^2-a_1^2\right) \sqrt{E^2-a_1^2}+E\left( a_2^2+a_1^2\right)
}<1$. If $r_1\rightarrow 2m$ then $\Vert v\Vert \rightarrow 1$.
\end{itemize}

An observer with $r>2m$ is unable to observe a free-falling
particle crossing the event horizon, since light rays cannot
escape from the zone $r\realleq 2m$. Hence, it can never observe a
free-falling particle reaching the speed of light. The only
observer being able to observe a particle at $r=2m$ is an observer
which crosses the event horizon at the same time and at the same
point as the particle. The relative velocity of the particle
observed by this observer is smaller than the speed of light, as
it is shown in \cite{Craw02}.

\subsection{Doppler effect and gravitational redshift}

\label{S3.2}

Taking into account Definition \ref{relvelu}, we can generalize
the expression of Doppler effect (\ref{f1.dop}) for observers at
different events of the same light ray:

\begin{proposition}
\label{cor.frcte}Let $\lambda $ be a light ray from $q$\ to $p$
and let $u$, $u'$\ be two observers at $p$, $q$\ respectively.
Then
\begin{equation}
\nu '=\gamma \left( 1-g\left( v,w\right) \right) \nu ,
\label{dopplergen}
\end{equation}
where $\nu $, $\nu '$ are the frequencies of $\lambda $ observed
by $u$, $u'$ respectively, $v$ is the relative velocity of $u'$
observed by $u$, $w$ is the relative velocity of $\lambda $
observed by $u$ and $\gamma $ is the gamma factor corresponding to
the velocity $\Vert v\Vert $.
\end{proposition}

\begin{proof} Let $F$ be the frequency vector field of $\lambda $. Then,
$\nu '=-g\left( F_q,u'\right) $. Since parallel transport
conserves metric, we have $\nu '=-g\left( \tau _{qp}F_q,\tau
_{qp}u'\right) =-g\left( F_p,\tau _{qp}u'\right) $. So, the
frequency of $\lambda $ observed by $\tau _{qp}u'$ is also $\nu
'$. Taking into account (\ref{f1.dop}) and Definition
\ref{relvelu}, expression (\ref{dopplergen}) holds.
\end{proof}

Note that the proof of Proposition \ref{cor.frcte} assures that
the frequency of $\lambda $ observed by $u'$ is the same as the
frequency of $\lambda $ observed by $\tau _{qp}u'$. Taking into
account Definition \ref{ucomoving}, if $u$ is comoving with $u'$
then they observe $\lambda $ with the same frequency. This result
can be also obtained from expression (\ref{dopplergen}), since the
relative velocity $v$ of $u'$ observed by $u$ is zero if $u$ is
comoving with $u'$.

\begin{figure}[tbp]
\begin{center}
\includegraphics[width=0.25\textwidth]{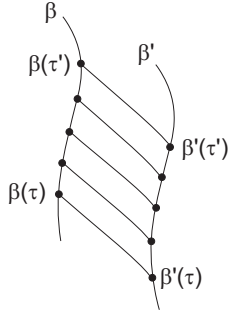}
\end{center}
\caption{If $\beta $ is comoving with $\beta '$ then $\beta $
observes that $\beta '$ uses the same clock as its.} \label{fig5}
\end{figure}

So, given $\beta $ an observer comoving with another observer
$\beta '$ and given $\lambda $ a light ray from $\beta '$ to
$\beta $, we have that $\beta $ and $\beta '$ observe $\lambda $
with the same frequency. Hence, within the framework of lightlike
simultaneity, ``{\it $\beta $ is comoving with $\beta '$}'' means
``{\it $\beta $ is spectroscopically comoving with $\beta '$}''.
This fact can be interpreted in this way: if $\beta '$ emits $n$
light rays in a unit of its proper time, then $\beta $ observes
also $n$ light rays in a unit of its proper time. So, $\beta $
observes that $\beta '$ uses the \textquotedblleft same
clock\textquotedblright\ as its (see Fig. \ref{fig5}).

Given two stationary observers (i.e. with constant spatial
coordinates, for a given coordinate system) $\beta $, $\beta '$,
and a light ray $\lambda $ from $\beta '$ to $\beta $, the
frequency of $\lambda $ observed by $\beta $ is, in general,
different from the frequency observed by $\beta '$. This
phenomenon is known as {\it gravitational redshift}. Since two
stationary observers are in \textquotedblleft
rest\textquotedblright\ with respect to each other, they are
supposed to be \textquotedblleft comoving\textquotedblright .
Thus, gravitational redshift effect has been always considered
independent from Doppler effect, arguing that photons lose or gain
energy when rising or falling in a gravitational field.
Nevertheless, in our formalism, stationary observers are not
comoving in general. Hence, there appears a Doppler shift given by
(\ref{dopplergen}) that coincides with the known gravitational
shift, explaining it in a natural way.

A clear example can be found in the Schwarzschild metric with
spherical coordinates, considering the radial light ray $\lambda $
given in (\ref{lightray}). Let $u_1$ be a stationary observer at
$q:=\lambda \left( r_1\right) $, and let $u_2$ be another
stationary observer at $p:=\lambda \left( r_2\right) $, with
$r_2>r_1>2m$. Taking $a_1:=a\left( r_1\right) $ and $a_2:=a\left(
r_2\right) $, we have that the relative velocity $v$ of $u_1$
observed by $u_2$ is given by (\ref{relvelschw}). Moreover, the
relative velocity $w$ of $\lambda $ observed by $u_2$ is
$a_2\left. \frac{\partial }{\partial r}\right\vert _{p}$. Applying
the general expression for Doppler effect (\ref{dopplergen}), we
have
\begin{equation}
\label{gravredschw} \nu _1=\frac{a_2 }{a_1 }\nu _2,
\end{equation}
where $\nu _1$, $\nu _2$ are the frequencies of $\lambda $
observed by $u_1$, $u_2$ respectively. This redshift is produced
because $u_2$ is not comoving with $u_1$ in our formalism.
Effectively, if we parallely transport $u_1$ to $p$ along $\lambda
$, we obtain the vector $\frac{1}{2}\left(
\frac{a_1}{a_2^2}+\frac{1}{a_1}\right) \left .\frac{\partial
}{\partial t}\right\vert _p+\frac{1}{2}\left( a_1
-\frac{a_2^2}{a_1}\right) \left. \frac{\partial }{\partial
r}\right\vert _p$, that it is obviously different from $u_2$.

Hence, we can affirm that given two equatorial stationary
observers $\beta _1\left( \tau \right) :=\left( \frac{1}{a_1}\tau
,r_1,\pi /2,0\right) $ and $\beta _2:=\left( \frac{1}{a_2}\tau
,r_2,\pi /2,0\right) $ with $\tau \in \mathbb{R}$, and a radial
light ray $\lambda $ from $\beta _1$ to $\beta _2$, equation
(\ref{gravredschw}) holds, where $\nu _1$, $\nu _2$ are the
frequencies of $\lambda $ observed by $\beta _1$, $\beta _2$
respectively. Equation (\ref{gravredschw}) is the known expression
for gravitational redshift in Schwarzschild metric, and so, it is
a particular case of the generalized Doppler effect given by
expression (\ref{dopplergen}). Note that $\nu \rightarrow 0$ when
$r_1\rightarrow 2m$, according to the fact that $\Vert v\Vert
\rightarrow 1$ when $r_1\rightarrow 2m$ (see (\ref{relvelschw})).

Another example is the cosmological redshift produced by the
expansion of the universe in the Robertson-Walker metric with
spherical coordinates $ds^{2}=-dt^{2}+a^{2}\left( t\right) \left(
\frac{1}{1-kr^{2}}dr^{2}+r^{2}\left( d\theta ^{2}+\sin ^{2}\theta
d\varphi ^{2}\right) \right) $, where $ a\left( t\right) $ is the
\textit{scale factor} and $k=-1,0,1$. Such redshift is too a
particular case of Doppler effect because stationary observers
(usually called \textquotedblleft comoving\textquotedblright ,
unfortunately for our formalism) are not comoving. This effect can
be calculated using the Killing $\left( 2,0\right) $-tensor
$K\left( X,Y\right) :=a^{2}\left( t\right) \left( g\left(
X,Y\right) +g\left( X,U\right) g\left( Y,U\right) \right) $ where
$X,Y$ are two vector fields and $U:=\frac{\partial }{\partial t}$
is the 4-velocity vector field of the congruence of stationary
observers. So, given $X$ a geodesic vector field, we have that
$K\left( X,X\right) =a^{2}\left( t\right) \left( g\left(
X,X\right) +g\left( X,U\right) ^{2}\right) $ is constant along its
integral curves. Therefore, since the frequency vector field $F$
of the light ray $\lambda $ is geodesic and lightlike, we have
that $a\left( t\right) g\left( F,U\right) $ is constant along
$\lambda $. So, $a\left( t\right) \nu $ is constant too, where
$\nu $ is the frequency of $\lambda $ observed by a stationary
observer of the congruence $U$. Hence, given two stationary
observers $\beta _1$, $\beta _2$ and a light ray $\lambda $
emitted by $\beta _1$ at coordinate time $t_1$ and observed by
$\beta _2$ at coordinate time $t_2$, we have that the expression
(\ref{dopplergen}) for Doppler effect has the form
\begin{equation}
\nu _1=\frac{a\left( t_2\right) }{a\left( t_1\right) }\nu _2,
\label{gravredrob}
\end{equation}
where $\nu _1$, $\nu _2$ are the frequencies of $\lambda $
observed by $\beta _1$ and $\beta _2$ respectively.

The functions $a\left( r\right) $ of (\ref{gravredschw}) and
$a\left( t\right) $ of (\ref{gravredrob}) are responsible for the
gravitational redshift in Schwarzschild and Robertson-Walker
metrics. This functions are usually called \textit{lapse
functions}.

\subsection{Light aberration}

\label{S3.3}

Taking into account Definition \ref{relvelu}, we can also
generalize expressions (\ref{f1.propor}) and (\ref{f1.cosabe}) of
light aberration effect for observers at different events of the
same light ray:

\begin{proposition}
\label{pro.abegen}Let $\lambda $ be a light ray from $q$\ to $p$
and let $u$, $u'$\ be two observers at $p$, $q$\ respectively.
Then
\begin{equation}
\tau _{qp}w'=\frac{1}{\gamma \left( 1-g\left( v,w\right) \right)
}\left( u+w\right) -\tau _{qp}u',  \label{abegen1}
\end{equation}
where $w$, $w'$ are the relative velocities of $\lambda $ observed
by $u$, $u'$ respectively, $v$ is the relative velocity of $u'$
observed by $u$, and $\gamma $ is the gamma factor corresponding
to the velocity $\Vert v\Vert $. Moreover, if $\tau _{qp}w'\neq w$
then
\begin{equation}
\cos \theta =\frac{\cos \theta '-\Vert v\Vert }{1-\Vert v\Vert
\cos \theta '},  \label{abegen2}
\end{equation}
where $\theta $ is the angle between $-w$ and $v$, and $\theta '$
is the angle between $-\tau _{qp}w'$ and the projection of $v$
onto $\left( \tau _{qp}u'\right) ^{\bot }$.
\end{proposition}

\begin{proof}
Let $F$ be the frequency vector field of $\lambda $. Then, $F_p
=\nu \left( u+w\right) $ and $F_q=\nu '\left( u'+w'\right) $.
Since $F$ is tangent to $\lambda $ and geodesic, we have $F_p=\tau
_{qp}F_q=\nu '\left( \tau _{qp}u'+\tau _{qp}w'\right) $. So, $\tau
_{qp}w'=\frac{\nu }{\nu '}\left( u+w\right) -\tau _{qp}u'$.
Applying Proposition \ref{cor.frcte}, expression (\ref{abegen1})
holds. If $\tau _{qp}w'\neq w$ then expression (\ref{abegen2}) is
obtained from (\ref{abegen1}) by simple algebraic manipulations.

\end{proof}

If $u$ is comoving with $u'$, then $\tau _{qp}u'=u$, $v=0$ and so,
from (\ref{abegen1}), we have $\tau _{qp}w'=w$. Since $\tau
_{qp}w'$ is the way $u$ observes $w'$, we can say that $u$ and
$u'$ observes $\lambda $ with the \textquotedblleft
same\textquotedblright\ relative velocity, and hence there is not
light aberration between comoving observers.

\section{Affine distance}

\label{S4}

\begin{figure}[tbp]
\begin{center}
\includegraphics[width=0.40\textwidth]{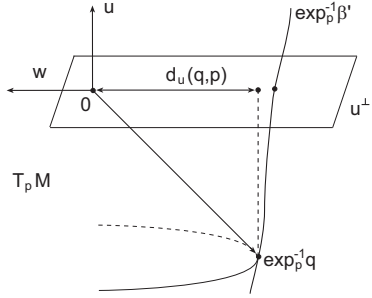}
\end{center}
\caption{Scheme in $T_{p}\mathcal{M}$ of the affine distance from
$q$ to $p$ observed by $u$, given in Definition \ref{d1.dist}. In
this case, $q$ is an event of a world line $\beta '$. Note that
$d_u\left( q,p\right) $ does not depend on $\beta '$.}
\label{fig3}
\end{figure}

To measure distances in our formalism we have to measure
\textquotedblleft lengths\textquotedblright\ of light rays, as we
told in the Introduction. But light rays are lightlike curves and
they have no length. To measure distances and angles, an observer
has to project these light rays onto its physical space (i.e. the
orthogonal space of its 4-velocity). This idea drives us to the
next definition of distance:

\begin{definition}
\label{d1.dist}Let $\lambda $ be a light ray from $q$\ to $p$ and
let $u$ be an observer at $p$. The affine distance from $q$ to $p$
observed by $u$, $d_u\left( q,p\right) $, is the module of the
projection of $\exp _{p}^{-1}q$ onto $u^{\bot }$ (see Fig.
\ref{fig3}).
\end{definition}

This concept of distance is defined according to the concept of
lightlike simultaneity given by the past-pointing horismos
submanifolds, because we measure distances between an event $p$
and the events that are observed simultaneously at $p$ (i.e.
events of $E_{p}^{-}$).

Taking into account Definition \ref{d1.dist}, we have $d_u\left(
q,p\right) =-g\left( \exp _{p}^{-1}q,w\right) $, where $w$ is the
relative velocity of $\lambda $ observed by $u$ (see Fig.
\ref{fig3}). So, it is easy to prove that
\begin{equation}
d_u\left( q,p\right) =g\left( \exp _{p}^{-1}q,u\right) .
\label{formdist}
\end{equation}
In the tangent space $T_{p}\mathcal{M}$ we have that $w$ and $\exp
_{p}^{-1}q $ are proportional and opposite. Taking into account
Definition \ref{d1.dist}, we have $\exp _{p}^{-1}q=-d_u\left(
q,p\right) \left( u+w\right) $. Given another observer $u'$ at
$p$, we have $\exp _{p}^{-1}q=-d_{u'}\left( q,p\right) \left(
u'+w'\right) $, where $w'$ is the relative velocity of $\lambda $
observed by $u'$. Therefore, we obtain
\begin{equation}
d_{u'}\left( q,p\right) =\gamma \left( 1-g\left( v,w\right)
\right) d_u\left( q,p\right) , \label{f1.dopdist}
\end{equation}
where $v$ is the relative velocity of $u'$ observed by $u$ and
$\gamma $ is the gamma factor corresponding to $\Vert v\Vert $.

If we compare (\ref{f1.dopdist}) with (\ref{f1.dop}), we realize
that frequency and affine distance have the same behaviour when a
change of observer is done. Hence, if $\lambda $ is a light ray
from $q$\ to $p$ and $u$, $u'$\ are two observers at $p$, we have
\begin{equation}
\frac{d_u\left( q,p\right) }{\nu }=\frac{d_{u'}\left( q,p\right)
}{\nu '}, \label{f1.concl}
\end{equation}
where $\nu $, $\nu '$ are the frequencies of $\lambda $ observed
by $u$, $u'$ respectively.

\begin{figure}[tbp]
\begin{center}
\includegraphics[width=0.35\textwidth]{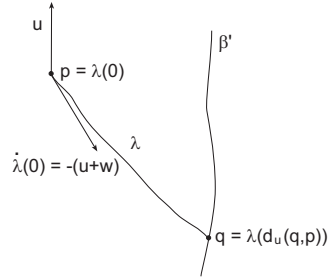}
\end{center}
\caption{Scheme of Proposition \ref{p2.pardis}, where $q$ is an
event of a world line $\beta '$. Note that $d_u\left( q,p\right) $
does not depend on $\beta '$.} \label{fig4}
\end{figure}

The next proposition shows that the concept of distance given in
Definition \ref{d1.dist} coincides with the known concept of
\textit{affine distance} introduced in \cite{kerm32}:

\begin{proposition}
\label{p2.pardis} Let $\lambda $ be a light ray from $q$ to $p$,
let $u$ be an observer at $p$ and let $w$ be the relative velocity
of $\lambda $ observed by $u$. If we parameterize $\lambda $
affinely (i.e. $\nabla _{\overset{.}{\lambda }\left( s\right)
}\overset{.}{\lambda }\left( s\right) =0 $) such that $\lambda
\left( 0\right) =p$, and $\overset{.}{\lambda }\left( 0\right)
=-\left( u+w\right) $, then $\lambda \left( d_u\left( q,p\right)
\right) =q$ (see Fig. \ref{fig4}).
\end{proposition}

\begin{proof}
By the properties of the exponential map (see \cite{Helg62}), we
have $\lambda \left( s\right) =\exp _{p}\left( -s\left( u+w\right)
\right) $. So $\lambda \left( d_u\left( q,p\right) \right) =\exp
_{p}\left( -d_u\left( q,p\right) \left( u+w\right) \right) =q$.
\end{proof}

Hence, given a light ray $\lambda $ from $q$ to $p$ and an
observer $u$ at $p$, we can interprete the affine distance from
$q$ to $p$ observed by $u$ as the distance (or time) travelled by
the light ray $\lambda $, measured by an observer at $p$ with
4-velocity $u$.

An equivalent result is given:

\begin{corollary}
\label{c2.pardis} Let $\lambda $ be a light ray from $q$ to $p$,
let $u$ be an observer at $p$ and let $w$ be the relative velocity
of $\lambda $ observed by $u$. If we parameterize $\lambda $
affinely such that $\lambda \left( 0\right) =q$, $\lambda \left(
d\right) =p$ and $\overset{.}{\lambda }\left( d\right) =u+w$, then
$d$ is the affine distance from $q$ to $p$ observed by $u$.
\end{corollary}

Now, we are going to generalize Definition \ref{d1.dist}:

\begin{definition}
Let $\beta $, $\beta '$ be two observers. The affine distance from
$\beta '$ to $\beta $ observed by $\beta $ is a real positive
function $d_{\beta }$ defined on $\beta $ such that, given $p\in
\beta $, $d_{\beta }\left( p\right) $ is the affine distance from
$q$ to $p$ observed by $u$, where $u$ is the 4-velocity of $\beta
$ at $p$, and $q$ is the unique event of $\beta '$ such that there
exists a light ray from $q$ to $p$.
\end{definition}

Note that even if $\beta $ is comoving with $\beta '$, the affine
distance $d_{\beta }$ from $\beta '$ to $\beta $ observed by
$\beta $ is not necessarily constant. Inversely, if $d_{\beta }$
is constant then $\beta $ is not necessarily comoving with $\beta
'$, as we will see in Section \ref{S4.2}. Only in some special
cases we have that $d_{\beta }$ is constant if and only if $\beta
$ is comoving with $\beta '$. For example in Minkowski if the
observers $\beta $ and $\beta '$ are geodesic.

Finally, we can define a distance on $E_p^-$ extending the concept
of affine distance given in Definition \ref{d1.dist}, using the
idea that an observer has to project light rays onto its physical
space:

\begin{definition}
\label{d2.dist}Let $u$ be an observer at $p$, and $q,q'\in
E_p^-\cup \left\{ p\right\} $. The affine distance from $q$ to
$q'$ observed by $u$, $d_u\left( q,q'\right) $, is the module of
$\pi _{u^{\bot }}\left( \exp _ p^{-1}q\right) -\pi _{u^{\bot
}}\left( \exp _p^{-1}q'\right) $, where $\pi _{u^{\bot }}$ is the
map ``projection onto $u^{\bot }$".
\end{definition}

It can be easily proved that
\begin{eqnarray}
d_{u}\left( q,q^{\prime }\right) &=&\left( g\left( \exp
_{p}^{-1}q-\exp _{p}^{-1}q^{\prime },\exp _{p}^{-1}q-\exp
_{p}^{-1}q^{\prime }\right)
\right.  \nonumber \\
& &\left. +g\left( u,\exp _{p}^{-1}q-\exp _{p}^{-1}q^{\prime
}\right) ^{2}\right) ^{1/2}. \label{distgen}
\end{eqnarray}
Moreover, expression (\ref{distgen}) generalizes expression
(\ref{formdist}) in the sense that if we substitute $q'$ by $p$ in
(\ref{distgen}), we obtain (\ref{formdist}).

The affine distance given in Definition \ref{d2.dist} is
symmetric, positive-definite and satisfies the triangular
inequality. So, it has all the properties that must verify a
topological distance defined on $E_p^-\cup \left\{ p\right\} $.

\section{Some examples of affine distance}

\label{S5}

In this Section we are going to show that affine distance is a
particular case of {\it radar distance} in the Minkowski
space-time (concretely, for geodesic observers), and generalizes
the {\it proper radial distance} in the Schwarzschild space-time.
Finally, we show that affine distance gives us a new concept of
distance in Robertson-Walker space-times, according to Hubble law.

\subsection{Minkowski}

In the Minkowski metric with rectangular coordinates
$ds^{2}=-dt^{2}+dx^{2}+dy^{2}+dz^{2}$, let us consider an event
$q=\left( t_{1},x_{1},y_{1},z_{1}\right) $ observed at $p=\left(
t_{2},x_{2},y_{2},z_{2}\right) $ by an observer $u=\gamma \left(
\left. \frac{\partial }{\partial t}\right\vert _{p}+v^{x}\left.
\frac{\partial }{\partial x}\right\vert _{p}+v^{y}\left.
\frac{\partial }{\partial y}\right\vert _{p}+v^{z}\left.
\frac{\partial }{\partial z}\right\vert _{p}\right) $, where
$\gamma $ is the gamma factor given by $\frac{1}{\sqrt{1-\left(
v^{x}\right) ^{2}-\left( v^{y}\right) ^{2}-\left( v^{z}\right)
^{2}}}$. Then, using (\ref{formdist}), we have the general
expression for the affine distance from $q$ to $p$ observed by
$u$:
\begin{eqnarray}
d_u\left( q,p\right) &=&g\left( q-p,u\right) \nonumber \\
&=&\gamma \left( \left( t_{2}-t_{1}\right) +v^{x}\left(
x_{1}-x_{2}\right) +v^{y}\left( y_{1}-y_{2}\right) +v^{z}\left(
z_{1}-z_{2}\right) \right) . \label{dmink1}
\end{eqnarray}
Note that $\left( t_{2}-t_{1}\right) =\sqrt{\left(
x_{1}-x_{2}\right) ^{2}+\left( y_{1}-y_{2}\right) ^{2}+\left(
z_{1}-z_{2}\right) ^{2}}$ because there is a light ray from $q$ to
$p$.

There exists a known method to measure distances between an
observer $\beta $ (that we can suppose parameterized by its proper
time $\tau $) and an observed event $q$, called \textquotedblleft
radar method\textquotedblright , consisting on emitting a light
ray from $\beta \left( \tau _{1}\right) $ to $q$, that bounces and
arrives at $p=\beta \left( \tau _{2}\right) $. The {\it radar
distance} between $\beta $ and $q$ observed by $\beta $ is given
by $\frac{1}{2}\left( \tau _{2}-\tau _{1}\right) $ \cite{Misn73}.
So, considering a geodesic observer $\beta $ passing through $p$
with 4-velocity $u=\gamma \left( \left. \frac{\partial }{\partial
t}\right\vert _{p}+v^{x}\left. \frac{\partial }{\partial
x}\right\vert _{p}+v^{y}\left. \frac{\partial }{\partial
y}\right\vert _{p}+v^{z}\left. \frac{\partial }{\partial
z}\right\vert _{p}\right) $ at $p$ we have that
\begin{eqnarray}
\beta \left( \tau \right) &=&\left( \gamma \left( \tau -\tau
_{2}\right) +t_{2},\gamma v^{x}\left( \tau -\tau _{2}\right)
+x_{2},\right.
\nonumber \\
& &\left. \gamma v^{y}\left( \tau -\tau _{2}\right) +y_{2},\gamma
v^{z}\left( \tau -\tau _{2}\right) +z_{2}\right) \label{betamink}
\end{eqnarray}
is the parameterization by its proper time. Setting out that
$q-\beta \left( \tau _{1}\right) $ is lightlike and $\tau
_{2}-\tau _{1}\neq 0$, from (\ref{betamink}) we obtain
\begin{eqnarray}
\frac{1}{2}\left( \tau _{2}-\tau _{1}\right) &=&\gamma \left(
\left( t_{2}-t_{1}\right) +v^{x}\left( x_{1}-x_{2}\right) \right.
\nonumber \\
& &\left. +v^{y}\left( y_{1}-y_{2}\right) +v^{z}\left(
z_{1}-z_{2}\right) \right) . \label{dmink2}
\end{eqnarray}
Comparing (\ref{dmink2}) with (\ref{dmink1}), we state that affine
distance coincides with radar distance for geodesic observers in
Minkowski space-time.

The radar distance between a non geodesic observer $\beta $ and an
observed event $q$ depends on the world line $\beta $ between
$\beta \left( \tau _{1}\right) $ and $\beta \left( \tau
_{2}\right) $. On the other hand, the affine distance only depends
on the 4-velocity of the observer at $p=\beta \left( \tau
_{2}\right) $, i.e. at the instant when the light ray arrives from
$q$. So, it is easier to calculate and it has more physical sense.

\subsection{Schwarzschild}

\label{S4.2}

In Schwarzschild metric with spherical coordinates, let $\beta
_{1}$ and $\beta _{2}$ be two stationary observers like in Section
\ref{S3.2}. We are going to calculate the affine distance $d$ from
$\beta _{1}$ to $\beta _{2}$ observed by $\beta _{2}$. Since
\[
\lambda \left( s\right) =\left( -a\left( r_{2}\right) s+2m\ln
\left( 1-\frac{s}{r_{2}\,a\left( r_{2}\right) }\right)
,r_{2}-a\left( r_{2}\right) s,\pi /2,0\right)
\]
is a light ray parameterized as in the hypotheses of Proposition
\ref{p2.pardis}, with $p:=\lambda \left( 0\right) \in \beta _2$
and $q:=\lambda \left( \frac{r_2-r_1}{a\left( r_2\right) }\right)
\in \beta _1$, we have that the affine distance from $q$ to $p$
observed by $u$ (where $u$ is the $4$-velocity of $\beta _2$ at
$p$) is given by $d_u\left( q,p\right) =\frac{r_2-r_1}{a\left(
r_2\right) }$. This expression only depends on $r_1$ and $r_2$,
i.e. the events $q$ and $p$ can be any events of $\beta _1$ and
$\beta _2$ respectively, such that there exists a light ray from
$q$ to $p$. Hence the affine distance $d$ from $\beta _1$ to
$\beta _2$ observed by $\beta _2$ is given by
\begin{equation}
d=\frac{r_2-r_1}{a\left( r_{2}\right) }. \label{dsch}
\end{equation}
So, $d$ is constant, but $\beta _{2}$ is not comoving with $\beta
_{1}$.

Expression (\ref{dsch}) is precisely a known expression for the
{\it proper radial distance} between spheres of radius $r_1$ and
$r_2$ (see \cite{Misn73}). So, the affine distance generalizes the
proper radial distance given in Schwarzschild metric.

\subsection{Robertson-Walker}

In Robertson-Walker metric with spherical coordinates, let $\beta
_{1}$ and $\beta _{0}$ be two stationary observers at $r=r_{1}>0$
and $r=0$ respectively. Let us suppose that $\beta _{1}$ emits a
light ray $\lambda $ at $t=t_{1}$ that arrives at $\beta _{0}$ at
$t=t_{0}$. To study distances in cosmology it is usual to consider
the scale factor in the form
\begin{equation}
a\left( t\right) =a\left( t_{0}\right) \left( 1+H_{0}\left(
t-t_{0}\right) -\frac{1}{2}q_{0}H_{0}^{2}\left( t-t_{0}\right)
^{2}\right) +\mathcal{O}\left( H_{0}^{3}\left( t-t_{0}\right)
^{3}\right)  \label{adet}
\end{equation}
where $a\left( t_{0}\right) >0$, $H\left( t\right)
=\overset{.}{a}\left( t\right) /a\left( t\right) $ is the Hubble
\textquotedblleft constant\textquotedblright , $H_{0}=H\left(
t_{0}\right) >0$, $q\left( t\right) =-a\left( t\right)
\overset{..}{a}\left( t\right) /\overset{.}{a}\left( t\right)
^{2}$ is the deceleration coefficient, and $q_{0}=q\left(
t_{0}\right) >0$, with $\left\vert H_{0}\left( t-t_{0}\right)
\right\vert \ll 1$ \cite{Misn73}. This corresponds to a universe
in decelerated expansion and the time scales that we are going to
use are relatively small.

The {\it proper distance}, $d_{proper}$, between two stationary
observers at a given instant $t$ is defined as the coordinate
distance multiplied by the scale factor $a\left( t\right) $ (see
\cite{Misn73}). The proper distance between $\beta _{1}$ and
$\beta _{0}$ at $t=t_{0}$ is given by $d_{proper}:=r_{1}a\left(
t_{0}\right) $. Obviously, this distance is not the same as the
affine distance (which we are going to denote $d_{affine}$). We
define the redshift parameter $z:=\frac{a\left( t_{0}\right)
}{a\left( t_{1}\right) }-1$, obtaining that%
\begin{equation}
d_{proper}=\frac{z}{H_{0}}\left( 1-\frac{1}{2}\left(
1+q_{0}\right) z\right) +\mathcal{O}\left( z^{3}\right) .
\label{dpresent_z}
\end{equation}

Moreover, the {\it luminosity distance}, $d_{luminosity}$, between
a stationary observer and a stationary light source at a given
instant $t$ is defined as $d_{luminosity}:=\sqrt{\frac{L}{4\pi
A}}$, where $L$ is the absolute luminosity and $A$ is the apparent
luminosity (see \cite{Misn73}). Applied to $\beta _{0}$ and $\beta
_{1}$ at $t=t_{0}$, we have
\begin{equation}
d_{luminosity}=\frac{z}{H_{0}}\left( 1+\frac{1}{2}\left(
1-q_{0}\right) z\right) +\mathcal{O}\left( z^{3}\right) .
\label{dlum_z}
\end{equation}
Comparing (\ref{dlum_z}) with (\ref{dpresent_z}), we obtain that
$d_{proper}<d_{luminosity} $ for $z\ll 1$. This distance is
related to the geodesic deviation method, and it is studied in
\cite{Newm59}.

Finally, we are going to calculate the affine distance
$d_{affine}$ from $\beta _{1}$ to $\beta _{0}$ observed by $\beta
_{0}$ at $t=t_{0}$. It can be interpreted as the distance
travelled by the light ray $\lambda $ measured by the observer
$\beta _{0}$, and it will satisfy $r_{1}a\left( t_{1}\right)
<d_{affine}<r_{1}a\left( t_{0}\right) =d_{proper}$. The vector
field $-\frac{1}{a}\frac{\partial }{\partial
t}+\frac{\sqrt{1-kr^{2}}}{a^{2}}\frac{\partial }{\partial r}$ is
geodesic, lightlike and its integral curves are radial light rays
that arrive at $r=0$ (i.e. at $\beta _{0}$). So, to parameterize
$\lambda $ like in Proposition \ref{p2.pardis}, we have to set out
the system
\begin{equation}
\left\{
\begin{array}{l}
\overset{.}{\lambda }^{t}\left( s\right) =\frac{-a\left(
t_{0}\right) }{a\left( \lambda ^{t}\left( s\right) \right) } \\
\overset{.}{\lambda }^{r}\left( s\right) =\frac{a\left(
t_{0}\right) \sqrt{1-k\lambda ^{r}\left( s\right)
^{2}}}{a^{2}\left( \lambda ^{t}\left(
s\right) \right) } \\
\lambda ^{t}\left( 0\right) =t_{0};~\lambda ^{r}\left( 0\right)
=0
\end{array}
\right. ,  \label{sistem}
\end{equation}
where $\lambda ^t$ and $\lambda ^r$ are the temporal and radial
components of $\lambda $ respectively. Using (\ref{adet}) and
taking into account that $\lambda ^{t}\left( d_{affine}\right)
=t_{1}$ (by Proposition \ref{p2.pardis}), from the integration of
the first equation of (\ref{sistem}) we obtain that
\begin{eqnarray}
\label{dlightlikerw} d_{affine} &=&\left( t_{0}-t_{1}\right)
-\frac{1}{2}H_{0}\left( t_{0}-t_{1}\right)
^{2}-\frac{1}{6}q_{0}H_{0}^{2}\left(
t_{0}-t_{1}\right) ^{3} \\
& &+\mathcal{O}\left( H_{0}^{3}\left( t_{0}-t_{1}\right)
^{3}\right) . \nonumber
\end{eqnarray}
Since $H_0\left( t_0-t_1\right) =z-\left(
1+\frac{1}{2}q_{0}\right) z^2+\mathcal{O}\left( z^3\right) $, from
(\ref{dlightlikerw}) we have
\begin{equation}
d_{affine}=\frac{z}{H_{0}}\left( 1-\frac{1}{2}\left(
3+q_{0}\right) z\right) +\mathcal{O}\left( z^{3}\right) ,
\label{dz}
\end{equation}
that is consistent with the Hubble law (for $z$ of first order
approximation). If we compare (\ref{dz}) with (\ref{dpresent_z})
we obtain that, effectively, $d_{affine}<d_{proper}$ for $z\ll 1$.

\end{document}